\documentclass[%
 aip,
 amsmath,amssymb,
 reprint,%
]{revtex4-1}

\usepackage{graphicx}
\usepackage{dcolumn}
\usepackage{bm}

\usepackage[utf8]{inputenc}
\usepackage[T1]{fontenc}
\usepackage{mathptmx}
\usepackage{etoolbox}
\usepackage{xcolor}

\DeclareUnicodeCharacter{00A0}{ }

\def\@email#1#2{%
 \endgroup
 \patchcmd{\titleblock@produce}
  {\frontmatter@RRAPformat}
  {\frontmatter@RRAPformat{\produce@RRAP{*#1\href{mailto:#2}{#2}}}\frontmatter@RRAPformat}
  {}{}
}%
\makeatother
\begin{document}

\preprint{AIP/123-QED}

\title[Combined Raman spectroscopy and electrical transport measurements in ultra-high vacuum down to 3.7~K]{Combined Raman spectroscopy and electrical transport measurements in ultra-high vacuum down to 3.7~K}
\author{K.P. Shchukin}
\email{konstantin.shchukin@tuwien.ac.at}
\affiliation{Institut f\"ur Festk\"orperelektronik, Technische Universit\"at Wien, Gu{\ss}hausstra{\ss}e 25, 1040 Vienna, Austria}
\affiliation{II. Physikalisches Institut, Universit\"at zu K\"oln, Z\"ulpicher Stra{\ss}e 77, 50937 Cologne, Germany}
\affiliation{Zentrum für Mikro- und Nanostrukturen, Technische Universit\"at Wien, Gu{\ss}hausstra{\ss}e 25, 1040 Vienna, Austria}
\author{M.~Hell}
\affiliation{II. Physikalisches Institut, Universit\"at zu K\"oln, Z\"ulpicher Stra{\ss}e 77, 50937 Cologne, Germany}
\author{A. Gr\"uneis}
\email{alexander.grueneis@tuwien.ac.at}
\affiliation{Institut f\"ur Festk\"orperelektronik, Technische Universit\"at Wien, Gu{\ss}hausstra{\ss}e 25, 1040 Vienna, Austria}

\date{\today}

\begin{abstract}
An instrument for the simultaneous characterization of thin films by Raman spectroscopy and electronic transport down to 3.7~K has been designed and built. This setup allows for the \textit{in-situ} preparation of air-sensitive samples, their spectroscopic characterization by Raman spectroscopy with different laser lines and five-probe electronic transport measurements using sample plates with prefabricated contacts. The lowest temperatures that can be achieved on the sample are directly proven by measuring the superconducting transition of a Niobium film. The temperature-dependent Raman shift and narrowing of the Silicon $F_{2g}$ Raman line are shown. This experimental system is specially designed for \textit{in-situ} functionalization, optical spectroscopic and electron transport investigation of thin films. It allows for easy on-the-fly change of samples without the need to warm up the cryomanipulator.
\end{abstract}

\maketitle

\section{Introduction}
\subsection{Ultra-high vacuum Raman spectroscopy}
Raman spectroscopy is a powerful and yet relatively easy experimental method that yields much more than the energies of Raman active vibrations from inelastically scattered light with an excellent energy resolution of about 1 wavenumber (0.125 meV). The Raman signal of many materials indirectly probes the electronic structure if it is resonant Raman scattering and is sensitive to doping and strain. The presented Raman setup is especially useful to probe chemically functionalized materials \textit{in situ} which allows for tuning a material's electronic properties. Many functionalization routes such as doping by foreign atoms are not air-stable and thus require ultra-high vacuum (UHV) conditions. Raman spectroscopy can be used to spectroscopically investigate chemical functionalization.  For example, the Raman spectrum contains valuable information on doping and strain for two-dimensional materials and organic thin films. Despite most Raman spectroscopy is carried out \textit{ex situ} in air, there are several reports on UHV Raman spectroscopy setups where functionalized carbon materials \cite{Hell2018}, III-V semiconductors \cite{Pierre23}, surface reconstructions and superstructures\cite{speiser2020vibrational} have been \textit{in situ} synthesized and measured. The setup presented 
 in this work advances the UHV Raman technique considerably by implementing also \textit{in situ} electronic transport which can be investigated simultaneously with the Raman spectroscopy. This setup can be applied to studying the change of electronic transport properties as a function of chemical functionalization.

\subsection{Low-temperature cryomanipulator}
For experiments in which \textit{in-situ} growth or chemical functionalization is performed, a low-temperature cryomanipulator with the possibility to exchange samples on the fly without the need to warm up the cryostat is desired. In this way, samples can be synthesized or chemically functionalized in the preparation chamber and transferred to the measurement chamber. To align the sample to the focus of the laser beam, the cryomanipulator requires translational ($x, y, z$) and polar ($\varphi$) movement. To date, related cryomanipulators with six axes ($x, y, z, \varphi$, azimuthal $\theta$ and axial tilt $\phi$ movements) are employed e.g. in angle-resolved photoemission spectroscopy (ARPES) measurements. In 2003, a goniometer with three rotating axes was designed which was based on the Janis-ST400 cryostat \cite{Aiura2003} and reached the lowest temperature of 12.5~K. The lowest temperature of 6.0~K on the sample insidea a six-axis cryomanipulator was reported in 2017. The temperature was proven by ARPES by the appearance of a quasi-gap at the Fermi level after cooling below critical temperature $T_c$ in FeSe superconductors~\cite{Hoesch2017}. In this case, the design was based on a Janis-ST402 cryostat. In 2020, a group reported a homebuilt flow cryostat that can cool down to a sample while rotation \cite{Tee2020}.

\subsection{Ultra-high vacuum electrical transport measurements}
We distinguish between three different experimental approaches for performing four-point transport measurements on a surface in UHV.
One approach is the use of prefabricated contacts on the sample. This approach has been used e.g. for the application of an electric field to a thin film in ARPES measurements \cite{Krempasky2018}. Since in this approach, the contacts are not fabricated \textit{in situ}, a cleaning procedure for the film is required after the sample has been brought into the UHV chamber. These cleaning procedures could involve e.g. annealing or sputtering the surface. In the above example, a thin Selenium film was used to cap the sample area to prevent exposure to ambient conditions. This Se could then be removed by annealing the sample in UHV. If one uses a sample holder and thermally stable contacts, this approach can also be used to prepare samples \textit{in situ}~\cite {Zhang2013}.
For the second approach, prefabricated microtips are used to establish electric contact with the sample through the motion of probe tips in UHV. This concept has been used by several works which probe e.g. the Si surface at room temperature \cite{Kanagawa2003,Hofmann2008,Perkins2013}. The low-temperature realization of this approach~\cite{Odobesco2010}, utilizing a closed-cycle cryostat, employed homemade molybdenum probes equipped with spring tips made from beryllium bronze and was capable of operating at temperatures as low as 22 K.
Finally, a third approach involves the deposition of metals onto the sample to be investigated to form good Ohmic contacts followed by contacting the contact pads with a tip\cite{Cui2020}.
In the present work, we will use approach one, in which the contacts are deposited onto the substrate before the insertion of the sample into the UHV chamber. This approach is especially suited for functionalizing samples stable in an inert atmosphere e.g. carbon materials. The chemical functionalization (e.g. by doping with alkali metals) is performed after the sample has been brought into UHV conditions.

\begin{figure*}[ht!]
    \centering
    \includegraphics[width=\textwidth]{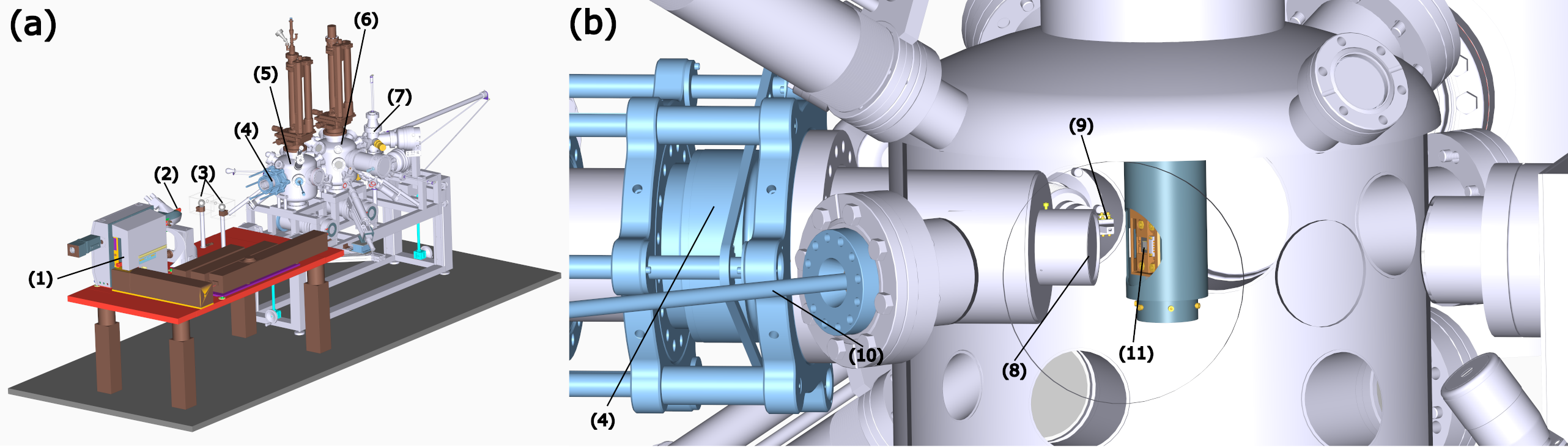}
    \caption{(a) Computer-aided design (CAD) drawing of the whole setup consisting of an optical table (in the foreground) onto which a Raman spectrometer (1) is placed. The laser beam is coupled out from a window (2) and directed by two mirrors (3) into the inverted optical flange (4) which is mounted on the analysis chamber (5). The sample synthesis is carried out in the preparation chamber (6) with attached sample storage (7). (b) The laser beam enters and excites the analysis chamber via a sapphire window mounted on the inverted optical flange (8) which separates the UHV environment from the air side.  The analysis chamber also features a wobble stick (9) for sample manipulation and a nut driver (10) for sample fixation on the manipulator. The sample (11) can be moved via the wobble stick in and out from the cryomanipulator.}
    \label{fig:setup}
\end{figure*}

\section{Experimental setup}
\subsection{Implementation of ultra-high vacuum Raman spectroscopy and electronic transport}
Figure~\ref{fig:setup} shows a sketch of the complete setup. Our setup consists of an ultra-high vacuum (UHV) system with preparation and analysis chambers and an optical table onto which a commercial Renishaw inVia\texttrademark~ spectrometer and three lasers are placed. 
The preparation chamber (base pressure $\sim5\times10^{-10}$mbar) (Fig.\ref{fig:setup}(6)) is equipped with metal and organic evaporators, an Argon sputter gun, a gas leak valve, a heater stage (temperatures up to 1500~K) and alkali-metal getters installed and can be used to grow samples onto sample plates with prefabricated contacts or to functionalize them chemically. Samples prepared in such a way can be transferred easily from the preparation into the analysis chamber (Fig.\ref{fig:setup}(5)) inside UHV. The Raman spectroscopy and electron transport measurements can be carried out inside the analysis chamber (base pressure $\sim5\times10^{-11}$mbar) with the sample inserted into the cryomanipulator. Also, the analysis chamber is equipped with alkali-metal getters to probe the sample electrically during its alkali-metal functionalization.

The spectrometer (Fig.\ref{fig:setup}(1)) is equipped with four laser lines between red and UV (633~nm, 532~nm, 488~nm and 325~nm) and optical components such as $\lambda/2$ and $\lambda/4$ plate, polarization filters etc. The laser beam is coupled from the Raman spectrometer (Fig.\ref{fig:setup}(2)) and directed via two mirrors (Fig.\ref{fig:setup}(3)) into an inverted flange (Fig.\ref{fig:setup}(4)) mounted on the analysis chamber. It can be focused onto the sample via a long working distance objective mounted inside the optical flange (Fig.\ref{fig:setup}(8)). During Raman and transport measurements, the sample can be transferred into the cryomanipulator (Fig.\ref{fig:setup}(11)) of the analysis chamber by the wobble stick (Fig.\ref{fig:setup}(9)) where it is tightly screwed with the socket wrench (Fig.\ref{fig:setup}(10)) onto the coldfinger of the cryostat.
Our choice of sample receptacle was the Omicron type which can be manipulated inside UHV via a wobble stick and is a well-established standard for other UHV techniques. We have extended this sample plate using five spring-loaded contacts and have used it in the past to characterize the electronic properties of the field-effect transistor (FET) based on alkali-metal-doped graphene nanoribbons \cite{senkovskiy2021tunneling}.

\begin{figure*}[bth!]
        \includegraphics[width=\linewidth]{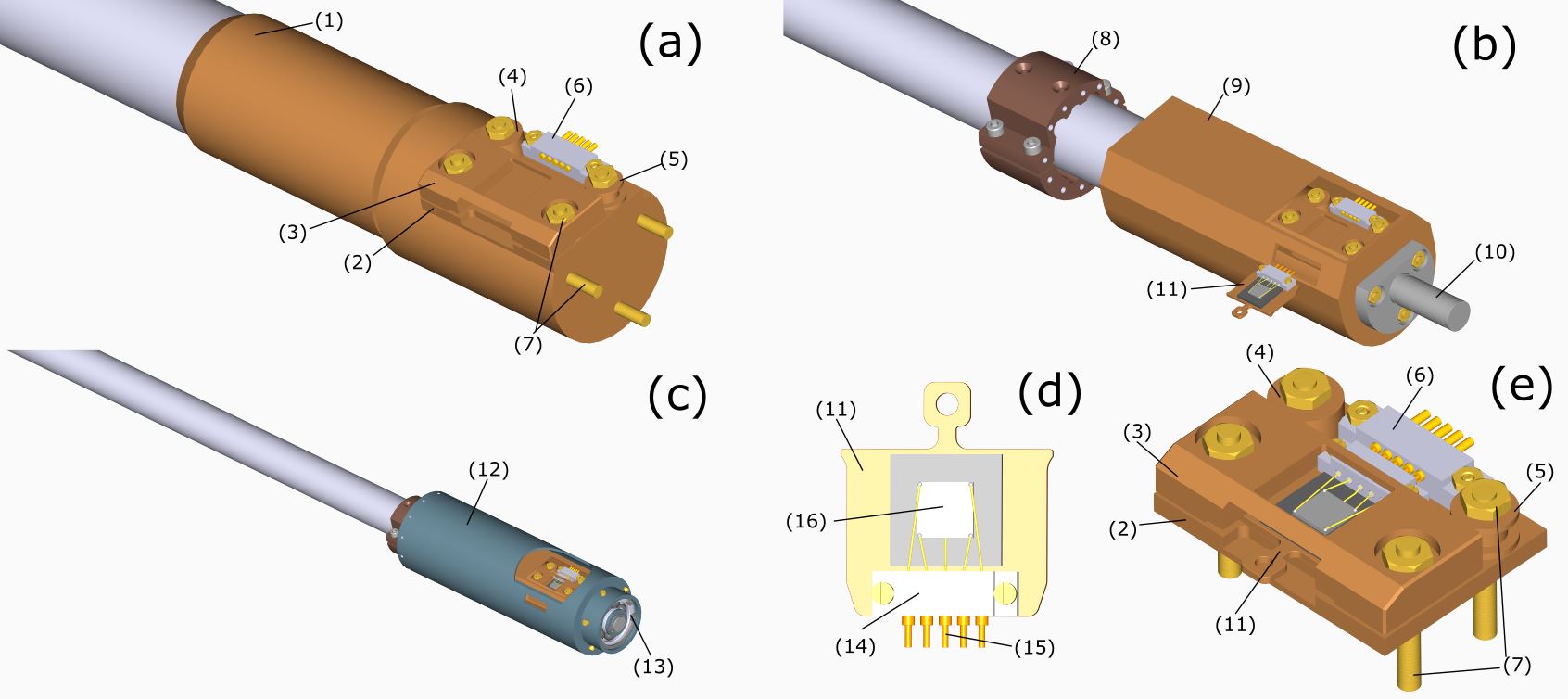}
        \caption{(a) A Janis ST-402 cryostat (1) was used as the basis for mounting a sample receptacle plate (2) with a top sample bracket (3). A RX-202A-CD (Rox\texttrademark) temperature sensor (4) and a heat sink bobbin (5) for measuring wires' thermalisation are mounted to a sample receptacle plate (2) in direct thermal contact. The counter Teflon block with five contacts (6) for the five spring contacts (15) and brass fasteners (7) for pressing the sample by the top bracket.      
        (b) The thermal anchor (8) for the rotatable outer shield (12) is mounted on a designated position of the cryostat. There is a fixed inner shield (9) in direct contact with the cold finger and a trunnion (10) for the rotatable outer shield. The sample plate with five spring-loaded contacts (11) can be inserted sideways.
        (c) The rotatable outer shield (12) covers much of the inner shield except an opening in front of the sample that allows for access to the screws needed for tightening the sample. The rotatable outer shield (12) is mechanically connected to the cryostat through the outer ring of a ceramic ball bearing (13). The inner ring of the bearing is mounted on a trunnion (10).
        (d) The sample plate is based on commonly used Omicron-type sample plates (11) with a Teflon housing (14) to hold five spring-loaded contacts (15) with leads attached to the sample (16) in a four-point geometry with back gating embedded in
        (e) Close-up of the sample receptacle with a sample plate loaded.}
        \label{fig:cryomanipulator}
\end{figure*}

\begin{figure}[htb!]
    \centering
    \includegraphics[width=\linewidth]{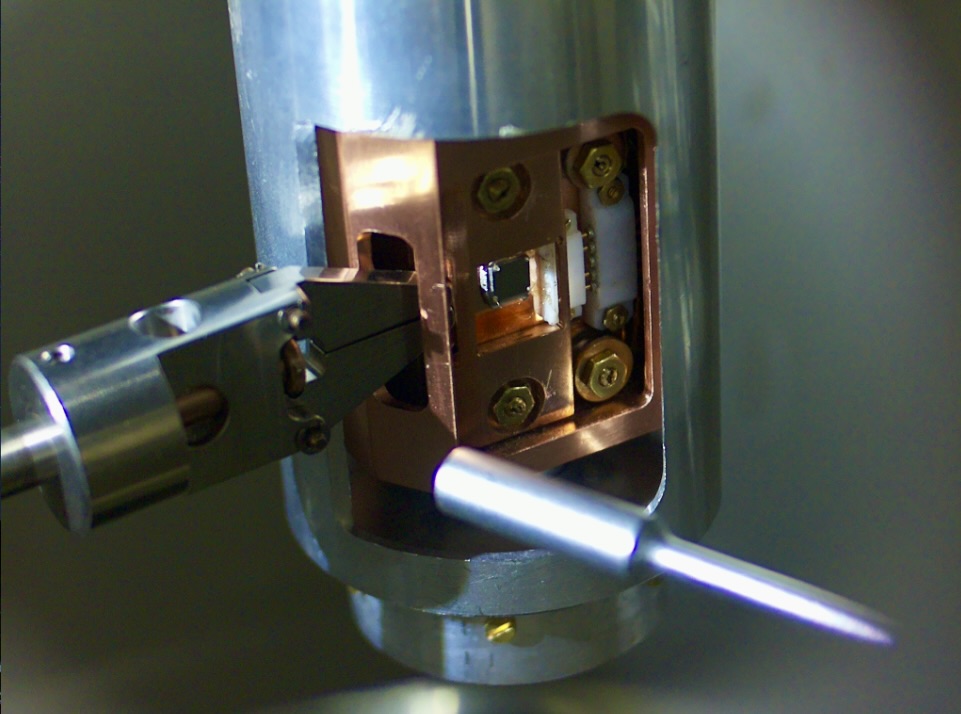}
    \caption{Photograph of a sample plate with Niobium film sputtered on a sapphire substrate. The sample plate is inserted into the receptacle with a wobble stick (from the left side). The socket wrench (from the right side) is used to tightly press the sample by the top bracket into the receptacle.
    }
    \label{fig:photo}
\end{figure}

\subsection{Cryomanipulator with optical access and electric contacts}
Figure \ref{fig:cryomanipulator} depicts a CAD drawing of the cryomanipulator construction. Janis ST-402 is used as the cryostat, which for general purposes includes liquid helium (lHe) transfer line with a porous plug, a Silicon diode temperature sensor in the cryostat's bath and the heater. The copper coldfinger of the cryostat (Fig.\ref{fig:cryomanipulator}(1)) has two milled platforms with threaded holes, which makes it possible to attach other parts to the coldfinger. Fastening the sample receptacle with brass nuts and threaded rods to the cold finger (Fig.\ref{fig:cryomanipulator}(7)) was used due to the lower thermal expansion of brass compared to copper. Thus thermal contact between fastened parts improves in low temperatures. A copper receptacle plate (Fig.\ref{fig:cryomanipulator}(2)) is flat polished with an average roughness of $R_a=0.13\mu$m. The low roughness is required for the interface maximisation between the plate and coldfinger. A top copper bracket (Fig.\ref{fig:cryomanipulator}(3)) is used for sample clamping and fastens with a brass nut on a brass threaded rod (Fig.\ref{fig:cryomanipulator}(7)). A Lake Shore RX-202A-CD temperature sensor (Fig.\ref{fig:cryomanipulator}(4)) was placed close to a sample place to minimize the temperature deviation between the measured temperature and the actual temperature of a sample. A heat sink bobbin (Fig.\ref{fig:cryomanipulator}(5)) out of oxygen-free copper was used as a thermal anchor for measuring wires to minimise heat transfer through them. All wires used for measuring electric transport are wrapped around the bobbin several times and fixed with a UHV-compatible silver conductive epoxy. Silver conductive epoxy has good heat conductivity and strong mechanical contact. The measuring wires are contacted to pin contacts installed in the Teflon block (Fig.\ref{fig:cryomanipulator}(6)) by one side, and another side is led to a feedthrough flange for connections to measuring units outside the vacuum system. Two shields protect a coldfinger and a sample receptacle from thermal radiation. An inner shield (Fig.\ref{fig:cryomanipulator}(9)) out of oxygen-free copper is strongly mechanically and thermally attached to a cold finger. An outer shield out of Aluminium (Fig.\ref{fig:cryomanipulator}(12)) is rotatable and thermally contacted to a thermal anchor (Fig.\ref{fig:cryomanipulator}(8)) by oxygen-free copper braids total crosssection of 25 mm$^2$. Mechanical connection to the cryostat of the rotatable outer shield is provided by a ceramic ball bearing (Fig.\ref{fig:cryomanipulator}(13)) and a trunnion (Fig.\ref{fig:cryomanipulator}(14)). Both shields have windows for sample transfer and for optical access to the sample.

Figure \ref{fig:photo} demonstrates the live view of sample loading and mounting on the cryostat. The manipulator position ($x, y, z, \varphi$) can be adjusted in such a manner that the sample plate is loaded into the receptacle in one forward movement of the wobble stick (from the left side). The final sample mounting with the socket wrench (from the right side) can be controlled from the same live view. During the tightening of the sample to the receptacle by the socket wrench, the sample is held by the wobble stick in order to compress the springs on the five pins that make electric contact. This establishes good electrical contact between spring-loaded pins on the sample plate and contact pins on the receptacle.

The cryostat cooling down from room temperature to 5K requires about 2l of lHe and takes about 40 min. The lHe consumption is about 1l/h to keep the cryostat temperature constant at about 5K.

\begin{figure*}[htb!]
    \centering
    \includegraphics[width=\textwidth]{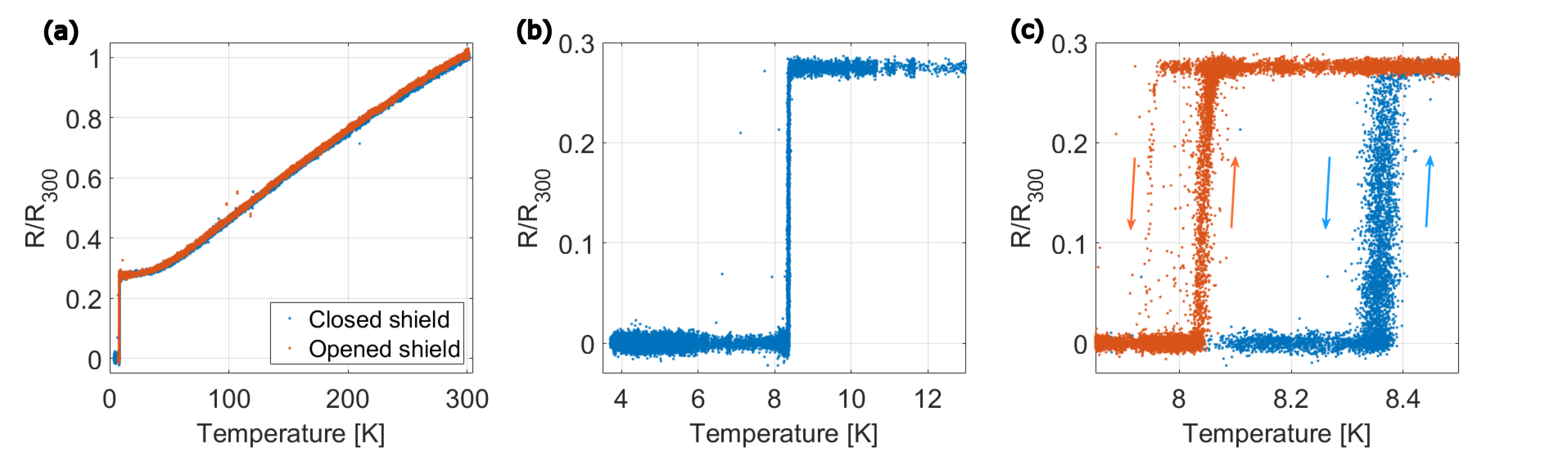}
    \caption{(a) Temperature dependence of a 100~nm Nb film on sapphire resistance normalised on room-temperature values $R_{300}$ with the outer shield opened and closed. Points are accumulated during cooling down and heating up. (b) Zoom-in of the same dependence measured with a closed shield demonstrates an obtained minimum temperature of 3.7~K. (c) Zoom-in of the same dependence measured with opened and closed shield demonstrates a sharp superconducting transition, arrows show the direction of temperature change. All measurements are carried out in ultra-high vacuum conditions.
    }
    \label{fig:fourpoint}
\end{figure*}

\subsection{Electronic transport measurements in ultra-high vacuum}
Five electrical wires (Manganin) have to be thermalized via a heat sink bobbin. The wires are fed through via the top of the manipulator and are braided to cancel out the induction electromotive forces (EMF) via external alternating magnetic flux. They are also shielded via a metal caging to protect the wires from external electric field induction. 

\begin{figure}[ht!]
    \centering
    \includegraphics[width=\linewidth]{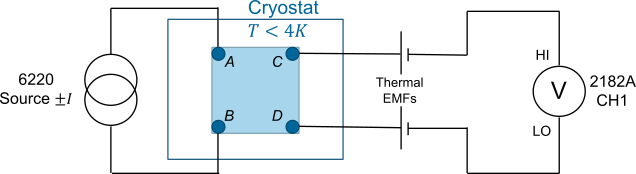}
    \caption{
    Circuit diagram of a thin metal-like film conductivity measurement with quasi-AC van der Pauw method.
    }
    \label{fig:deltascheme}
\end{figure}

The contact configuration on the feedthrough and the sample plate allows the application of different measuring units and sample geometries well established for 2D materials. Here we describe the four-point measurement configuration we use to study electron transport in metal-like thin films. The four-point measurements were provided every $1$~sec via a quasi-AC van der Pauw measurement technique which alters the current polarity (Figure~\ref{fig:deltascheme}). In particular, a precision current source Keithley 6220 creates the closed current loop with the sample through the contacts $A$ and $B$ and a nanovoltmeter Keithley 2182A  measures induced voltage drop between contacts $C$ and $D$. This technique is needed to remove thermal EMFs that develop during cooling across the contacts of two different metals along the whole line. 
For the resistance measurement, a current $I_{AB}$ was injected to the contact $A$ and taken out of contact $B$ and a voltage drop $V_{CD}(I_{AB})$ between contacts $C$ and $D$ was measured. Then, for the reversed polarity current $I_{BA}=-I_{AB}$ was applied and a voltage $V_{CD}(I_{BA})$ was obtained. If a thermal EMF is induced in the voltmeter measuring loop, it can be eliminated by taking the difference between the two measurements. This works because the thermal EMF does not change its polarity when you reverse the polarity of the measuring current. The resistance is then obtained via
\begin{equation}
  R = \frac{\vert V_{CD}(I_{AB}) - V_{CD}(I_{BA})\vert}{2I_{AB}}.  
\end{equation}

\section{Performance tests}
\subsection{Cooling performance and temperature stability}
Generally, in many comparable experimental setups, achieving the lowest possible temperature on the sample plate is a real challenge. The manufacturer specifies that the cold finger of the cryostat can go down to 4.2~K in regular flow and to about 2.2~K for a short time when pumping on the lHe reservoir. However, it is challenging to have the lowest temperatures on the sample. This is because the sample should be transferable and thus cannot be glued directly onto the coldfinger. Therefore, when using transferable samples, the precise determination of the sample temperature is crucial. In a previously built cryomanipulator, the temperature was determined using photoemission spectroscopy via the quasi-gap opening of a superconducting film\cite{Hoesch2017}. In the present case, the design of our cryomanipulator already provides the electrical contacts needed for the four-point electron transport measurement. That allows us to perform direct temperature-dependent resistance measurement of the Niobium (Nb) sample and thus define its critical temperature. Bulk Nb has a known critical temperature of $T_c=9.28~K$, which decreases in thin film samples due to reduced dimensionality \cite{Asada1969}.

Figure \ref{fig:fourpoint}~(a) depicts the four-point measurements of an RF-sputtered Nb film grown on sapphire\cite{Saito1975}. Four contacts to the film were made \textit{ex situ} and the sapphire was mounted on a sample plate with spring contacts and loaded into the UHV system, where it was annealed before cooling down. The sample resistance was measured with the shield in closed and opened positions. The resistance decreases from its room temperature value and suddenly drops to zero at a temperature of about 8~K. Below this temperature, no deviation of resistance was observed in temperatures down to $T_{min}=3.7$~K (Fig. \ref{fig:fourpoint}~(b)) in the case of the closed shield. More data points around the transition temperature were accumulated by heating up and cooling down the sample in the temperature range of $T=7.8\div8.5$~K. A zoom-in view shown in Figure \ref{fig:fourpoint}(c) to the region of the sudden drop reveals the temperature of the step at $T_c=8.36\pm0.03$~K for closed shield and  $T_c=8.00\pm0.06$~K for open shield.

The temperature difference between the closed and open shield conditions can be explained as follows. When the shield is open, the receptacle must be cooled more to compensate for the influx of photons. Since the temperature measurement is taken from the Rox\texttrademark  sensor close to the sample receptacle, the receptacle must reach a slightly lower temperature with the open shield compared to the closed shield scenario. The transition temperature is independent of whether the shield is open or closed. Therefore, we attribute the slightly lower transition temperature with the open shield to the increased cooling required to counteract the additional heat radiation. The presence of temperature hysteresis supports this explanation and can be attributed to the higher cooling power of lHe compared to the adjustable and lower heating power of the heater. 

\subsection{Ultra-high vacuum Raman spectroscopy of Silicon}
\begin{figure}[ht!]
    \centering
    \includegraphics[width=7cm]{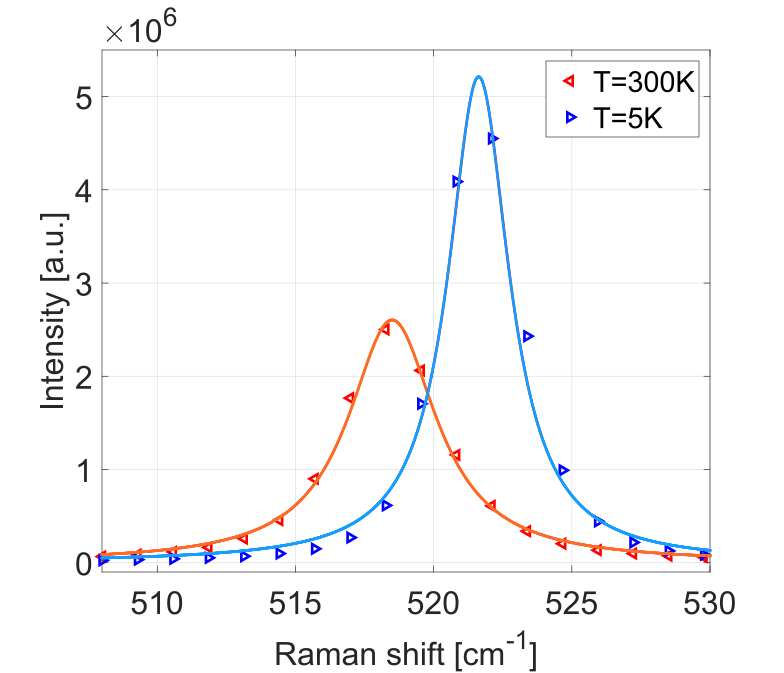}
    \caption{Raman spectra of a silicon wafer at 300~K and at 5~K measured with 532~nm laser with a 5~mW laser power. The spectra are recorded in the ultra-high vacuum Raman system shown in Figure 1.}
    \label{fig:raman}
\end{figure}

A Silicon wafer (dimensions 6 mm$\times$ 10 mm) was mounted onto a sample receptacle, brought into the UHV chamber and thoroughly outgassed. Raman spectra of the Si $F_{2g}$ Raman mode at around 520 cm$^{-1}$ were recorded inside the analysis chamber at room temperature (300~K) and the temperature of 5~K. In Figure~\ref{fig:raman}, the Raman spectra are shown. An energy shift of the Raman peak by $\vert\Delta\omega_{F_{2g}}\vert=3.1$~cm$^{-1}$ from 5~K to 300~K  can be observed. This is a consequence of the lattice expansion $\Delta V$. The $\Delta V$ is calculated from previously observed lattice parameters (see Table~\ref{tab:table1}).
Furthermore, we observe a narrowing of the linewidth when going from 300 K to 5 K. This narrowing of the Raman linewidth is explained as follows. The Raman linewidth is inversely proportional to the phonon lifetime which depends on the various scattering mechanisms (electron-phonon, phonon-phonon and defect scattering) and the experimental broadening. When temperature is decreased, the phonon scattering becomes weaker and hence the lifetime is longer and the Raman peak is sharper. 

Using this experimental data and previously published results on the temperature dependence of the Si lattice constants~\cite{Reeber1996259} (Table~\ref{tab:table1}), we can evaluate the Grüneisen parameter for this Si wafer. Grüneisen parameter of Si $F_{2g}$ mode can be expressed as 
\begin{equation}
    \gamma_{F_{2g}}=-\frac{V}{\omega_{F_{2g}}}\cdot\frac{\Delta\omega_{F_{2g}}}{\Delta V},
\end{equation}
where $\omega_{F_{2g}}$ and $V$ are $F_{2g}$ phonon mode frequency and the volume of the unit cell respectively. The relative volume change $\frac{\Delta V}{V}$ for cubic lattice in terms of lattice constant $a$ and its change with temperature $\Delta a$ can be approximated to $\frac{\Delta V}{V}\approx3\frac{\Delta a}{a}$. This yields Grüneisen parameter $\gamma_{F_{2g}}=0.23$, which is consistent with previous results~\cite{gauster1971low}.

\begin{table}
\caption{\label{tab:table1}Experimental values of Si $F_{2g}$ Raman mode $\omega_{F_{2g}}$ and linear expansion values~\cite{Reeber1996259} $a$ of Si cubic lattice for different temperatures.}
\begin{ruledtabular}
\begin{tabular}{c c c}
    Temperature,~K & $\omega_{F_{2g}}$,~cm$^{-1}$ & $a$,~\AA \\
\hline
5 & 521.63 & 5.429820\\
300 & 518.50 & 5.431092\\
\end{tabular}
\end{ruledtabular}
\end{table}

\section{Conclusions and Discussions}
In conclusion, we have demonstrated a setup for combined UHV Raman and electrical transport measurements that operates reliably down to sample temperatures of 3.7~K. Our setup allows for the easy and fast exchange of samples without having to stop the lHe cooling. The combination of Raman spectroscopy and electrical transport is particularly suitable for \textit{in-situ} sample functionalization such as chemical doping, where the sample has to be transferred frequently between the analysis and the preparation chambers. This setup is useful to tackle e.g. the following problems.

The investigation of alkali metal-doped carbon-based superconductors: it is well known that bulk C$_{\rm 60}$ becomes superconducting upon alkali metal doping. Yet, for thin films, experiments are scarce and to date only one report on potassium doped K$_3$C$_{\rm 60}$ films exists~\cite{Rogge2003}. In the bulk, Rubidium doped C$_{\rm 60}$, Rb$_3$C$_{\rm 60}$ has a higher critical temperature than K$_3$C$_{\rm 60}$. For thin films, the substrate effect is stronger and the two-dimensionality affects the critical temperature. It is therefore interesting to investigate Rb$_3$C$_{\rm 60}$ layers by a UHV-Raman and electrical transport combination. The UHV Raman technique is crucial for this problem because Rb doped C$_{\rm 60}$ is air sensitive and the different phases (superconducting Rb$_3$C$_{\rm 60}$ and insulating Rb$_6$C$_{\rm 60}$) have a distinct Raman response. Similarly, this setup is useful for discovering new, alkali-metal-doped organic superconductors.

Additionally, the system can be useful in investigating charge-density-wave (CDW) systems. There is a sudden change in the electrical properties across a CDW transition, often accompanied by the appearance of new Raman modes due to the back folding of the phonon dispersion relation.

The presented experimental setup can be used to study hydrogen or deuterium (H/D) functionalization experiments of 2D materials and surfaces. E.g. when graphene is exposed to a beam of atomic H/D, the graphene readily transforms from sp$^2$ to sp$^3$ bonding which induces scattering centers and a distinct change in the Raman D/G ratio. The evolution from purely sp$^2$ graphene to graphene with a large concentration of sp$^3$ defects could be investigated in this way in-situ.

Recently, the gap opening in double-side hydrogenated free-standing graphene was demonstrated\cite{betti2022gap}. A similar effect was theoretically predicted\cite{pujari2011single} for single-side hydrogenated graphene, but up to today has not been experimentally realised. Our setup could be used to perform the single-sided hydrogenation of graphene in-situ and characterize the sample by Raman and transport.

Some materials require low-temperature synthesis routes which can also be performed in the present setup. For instance, Lithium superoxide LiO$_2$ (a material that is relevant for its magnetic properties) requires matrix isolation techniques at temperatures of $15\div40$K for its synthesis. Lithium superoxide could be synthesized in the present setup as a single film and can immediately be pre-characterized by Raman spectroscopy and electronic transport.

This experimental setup can also be useful for the synthesis, Raman spectroscopic pre-characterisation and electronic transport study by measuring differential conductivity proportional to the density of stated (DOS) close to the Fermi level with the Fermi level variation by back gating.

\section{Acknowledgements}
The authors are grateful to W. Schrenk for help in Nb film magnetron sputtering. K.P.S. acknowledges I. A. Cohn for useful discussions and suggestions on applied cryogenics. A.G. acknowledges DFG project GR3708/4 and the FFG (Project CrystalGate).

\section{References}
\bibliography{papers}

\end{document}